# comfort-db: A Python toolkit to access, preprocess and explore the oceanographic COMFORT dataset

Yvonne Jenniges[1,2], Are Olsen[3], Boris Peter Koch[2,4], Sebastian Maneth[1]

06 November 2026

[1] University of Bremen, Bibliothekstraße 1, Bremen 28359, Germany

[2] Alfred-Wegener-Institut, Helmholtz-Zentrum für Polar- und Meeresforschung, Am Handelshafen 12, Bremerhaven 27570, Germany

[3] University of Bergen, Postboks 7800, Bergen 5020, Norway

[4] University of Applied Sciences, An der Karlstadt 8, Bremerhaven 27568, Germany

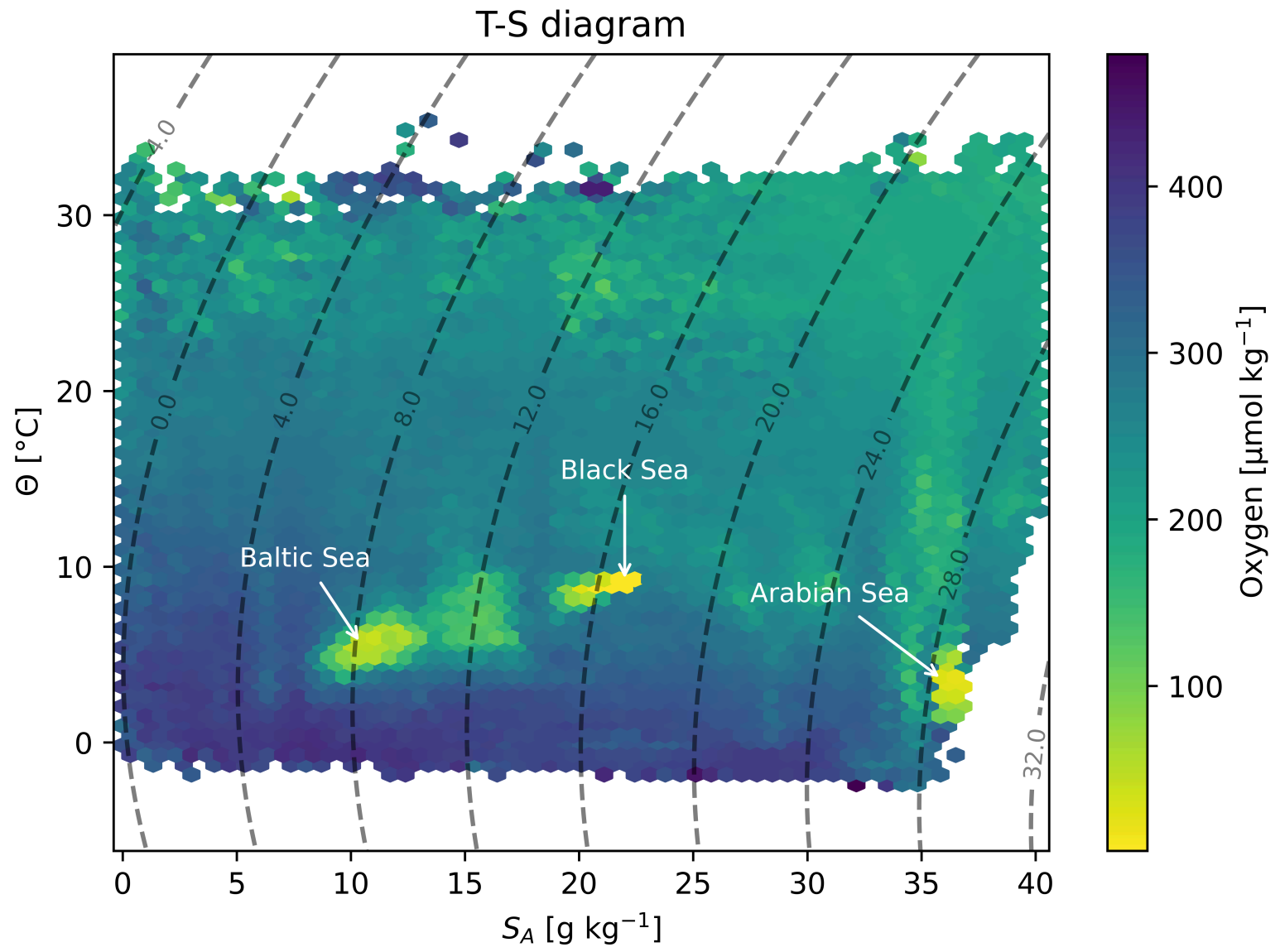

*Figure 1: Global temperature-salinity diagram (`comfort.plot`) with isopycnals in kg m$^{-3}$ (dashed lines, reference pressure 0 dbar, computed as density-anomaly) of oxygen measurements from the COMFORT dataset. Only quality-controlled (`QC_GOOD`) data are loaded, units are harmonised (`comfort.units`) and oxygen values that are identical to their co-located practical salinity are dropped (`flag_salinity_like_oxygen`). Duplicate spatio-temporal locations are averaged (`average_duplicates_by_location`) and temperature, salinity and oxygen inner-joined. Absolute Salinity ($S_A$) and Conservative Temperature ($\Theta$) are derived via TEOS-10 (`comfort.physics`). The remaining 40,434,271 data points (out of 67,085,418 oxygen observations in the dataset) are averaged into hexagonal bins and coloured by oxygen concentration. The three visible low-oxygen zones in the plot broadly correspond to known oxygen-depleted areas in the Baltic, Black and Arabian Sea (Paulmier & Ruiz-Pino, 2009).*

# Summary

`comfort-db` is a Python library that facilitates working with the COMFORT SQLite dataset, version 2 (Korablev et al., 2022). The dataset assembles global oceanographic measurements from ten datasets and contains a total of 458,724,734 in situ measurements from the years 1772 to 2020. `comfort-db` enables easy access to the database and provides preprocessing and exploratory analysis utilities. Preprocessing supports variable scaling and filtering by space, time and quality. Unit conversions are implemented as described in the COMFORT compilation report (Korablev et al., 2021) and can be applied during data loading or separately via a converter object. TEOS-10 physical oceanography (`comfort.physics`) is built on the `gsw` library (McDougall & Barker, 2011). Gridding is supported and optionally works with bathymetry data (e.g. GEBCO (GEBCO Compilation Group, 2022)) to exclude grid cells on land. Moreover, the library supports general and field-specific exploratory data analysis through plotting functions, an interactive dash app and functions to compute oceanographic profile and water mass statistics.

`comfort-db` builds on and extends code originally developed for (Jenniges et al., 2025), including parts of the gridding, unit conversion, database and plotting routines.

# Statement of need

Currently, a large preprocessing pipeline is required to work with the COMFORT dataset. Users must perform SQLite queries, interpret quality flags and convert units, which involves computing TEOS-10 variables and averaging temperature and salinity variables to get a unique match for an oxygen observation. Depending on the use case, users may also want to interpolate (depth) profiles, perform scaling and gridding. `comfort-db` integrates these tasks into one reproducible Python workflow and thus contributes to sustainable and transparent analyses built on the COMFORT dataset. It enables users of the COMFORT dataset, oceanographers without a strong software background and data scientists without a strong oceanographic background, to focus on scientific questions rather than repeated, error-prone data preprocessing and wrangling.

# State of the field

A related library is `argopy` (Maze & Balem, 2020) that enables data access, manipulation and visualisation of Argo float data. However, the COMFORT dataset differs from Argo not only in file format (SQLite vs. NetCDF), but also in scope and structure: It merges ten datasets into a station/profile hierarchy with its own SQLite schema, quality flag conventions and unit definitions and intended conversions. A simple adapter to `argopy` would not cover this schema- and domain-specific functionality. To our knowledge, no existing package offers a comparable reproducible interface to the COMFORT dataset.

# Software design

The organisation of the `comfort-db` library is driven by usability and the scale of the COMFORT dataset. Since the dataset is distributed as a single multi-gigabyte SQLite file, efficiency and low memory usage were key design goals. For example, space, time and quality filtering is applied at the SQL level rather than after loading potentially large data tables into memory. SQLite does not support spatial (polygon-based) filtering, e.g. to extract a specific ocean basin. Therefore, `comfort-db` first computes a broader bounding box, filters for it in the database and subsequently filters the data in-memory to the exact polygon shape. Moreover, `distance_to_coast` and `classify_from_grid` map coordinates to the unit sphere and build a k-d tree (`scipy`, (Virtanen et al., 2020)) to efficiently query the nearest coastline point or grid cell.

Quality flag interpretation and unit conversions are centralised in dedicated modules (`comfort.qc`, `comfort.units`) so that corrections and settings apply everywhere consistently. To cater to a diverse audience, more specialised functionality such as plotting, geospatial analysis, gridding, scaling and the dash app is available through optional installs (`plot`, `geo`, `grid`, `scale`, `dash`). Thus, the core package stays lightweight for users who only need database access and preprocessing.

Some processing steps require a single parameter value per spatio-temporal location, which is not guaranteed in the COMFORT dataset. For example, there can be different instruments and multiple bottles on a conductivity-temperature-depth (CTD) rosette at the same time, depth and coordinates. Also, there are measurements that were contributed by more than one of the merged source datasets under different station IDs. Therefore, `average_duplicates_by_location` averages the records of each parameter by longitude, latitude, depth and time, irrespective of station ID, profile, instrument or bottle number. Records that the dataset itself flagged as duplicates (PQF2=-1) can be removed by applying the `QC_GOOD` quality filter. When auxiliary temperature and salinity are added to another parameter table for unit conversions, they are first averaged by station and depth (in `load_comfort`), which likewise averages all profiles, instruments and bottles of a station into one representative value.

Figure 1 was generated by example 7 of the codebase, which illustrates quality filtering, TEOS-10 conversion and unit harmonisation. The data were then rendered as a global temperature-salinity diagram coloured by oxygen concentration.

# Research impact statement

The COMFORT dataset provides a large compilation of oceanographic observations, including data from established resources such as the World Ocean Database (Boyer et al., 2018) and the Global Ocean Data Analysis Project (Olsen et al., 2016, 2019). Accessing, preprocessing and exploring this data can require substantial technical effort and `comfort-db` lowers the barrier to exploit this extensive resource to facilitate future analyses and reproducible workflows.

To this point, (Jenniges et al., 2025) is the only published study using the COMFORT dataset and investigated three-dimensional biogeochemical provinces in the North Atlantic. Future analyses (Jenniges, in prep.) will apply the data to spatio-temporal imputation. The preprocessing of both studies can be closely reproduced using `comfort-db` (examples 8 and 9, respectively), with minor numerical differences from refinements to the density conventions since the original analyses.

# AI usage disclosure

Generative AI (Anthropic Claude Code, primarily Claude Sonnet 5) was used under direct supervision of the corresponding author (Y.J.) to restructure the author's own pre-existing analysis scripts into an installable package, update accompanying documentation, draft the majority of the tests and assist in visualisations and identifying and fixing bugs. All code was reviewed by the corresponding author, who implemented the original scientific logic. Correctness was assessed via the tests, checks against manually computed unit conversions (`CONVERSION_REFERENCE` in `test/test_units.py`) and execution of all example scripts.

# Acknowledgements

The first author (Y.J.) has been funded by the University of Bremen, the Helmholtz School for Marine Data Science (MarDATA) and the Alfred-Wegener-Institut, Helmholtz-Zentrum für Polar- und Meeresforschung, which also provided the computing infrastructure for this work.